\documentclass[%
    reprint,
    superscriptaddress,
    amsmath,
    amssymb,
    longbibliography,
    aps,
]{revtex4-1}
\usepackage{graphicx}
\usepackage{braket}
\usepackage{mathtools}
\usepackage{hyperref}
\hypersetup{
    colorlinks=true,
    linkcolor=blue,
    citecolor=cyan,
    urlcolor=cyan,
    }

\normalsize 
\usepackage{array} 
\usepackage{booktabs} 

\usepackage{acronym}
\newacro{LHC}{Large Hadron Collider}
\newacro{HEP}{High Energy Physics}
\newacro{QML}{Quantum Machine Learning}
\newacro{ML}{Machine Learning}
\newacro{SM}{Standard Model}
\newacro{BSM}{Beyond Standard Model}
\newacro{MC}{Monte Carlo}
\newacro{AD}{Anomaly Detection}
\newacro{AE}{auto-encoder}
\newacro{QSVM}{Quantum Support Vector Machines}
\newacro{NISQ}{Noisy Intermediate Scale Quantum}
\newacro{TPR}{True Positive Rate}
\newacro{FPR}{False Positive Rate}
\newacro{PCA}{Principal Component Analysis}
\newacro{VQC}{Variational Quantum Circuit}
\newacro{MSE}{Mean Squared Error}
\newacro{ROC}{Receiver Operating Curve}
\newacro{AUC}{Area Under Curve}
\newacro{GQC}{Guided Quantum Compression}

\newcommand{\tth}{$t\bar{t}H(b\bar{b})$}
\newcommand{\ttbb}{$t\bar{t}(b\bar{b})$}

\begin{document}
\title{Guided Quantum Compression for High Dimensional Data Classification}

\author{Vasilis Belis}
\thanks{Corresponding author: \href{mailto:vbelis@phys.ethz.ch}{vbelis@phys.ethz.ch}}
\affiliation{Institute for Particle Physics and Astrophysics, ETH Zurich, 8093 Zurich, Switzerland}
\author{Patrick Odagiu}
\affiliation{Institute for Particle Physics and Astrophysics, ETH Zurich, 8093 Zurich, Switzerland}
\author{Michele Grossi}
\affiliation{European Organization for Nuclear Research (CERN), CH-1211 Geneva, Switzerland}
\author{Florentin Reiter}
\affiliation{Institute for Quantum Electronics, ETH Zurich, 8093 Zurich, Switzerland}
\author{G\"unther Dissertori}
\affiliation{Institute for Particle Physics and Astrophysics, ETH Zurich, 8093 Zurich, Switzerland}
\author{Sofia Vallecorsa}
\affiliation{European Organization for Nuclear Research (CERN), CH-1211 Geneva, Switzerland}

\date{\today}

\begin{abstract}
Quantum machine learning provides a fundamentally different approach to analyzing data. 
\mbox{However},~many interesting datasets are too complex for currently available quantum computers. 
Present quantum machine learning applications usually diminish this complexity by reducing the dimensionality of the data, e.g., via auto-encoders, before passing it through the quantum \mbox{models}. 
Here, we design a classical-quantum paradigm that unifies the dimensionality reduction task with a quantum classification model into a single architecture: the guided quantum compression model.
We~exemplify how this architecture outperforms conventional quantum machine learning approaches on a challenging binary classification problem: identifying the Higgs boson in proton-proton collisions at the LHC.
Furthermore, the guided quantum compression model shows better performance compared to the deep learning benchmark when using solely the kinematic variables in our dataset.
\end{abstract}

\maketitle


\section{Introduction}\label{sec:intro}
\ac{ML} is established as an invaluable tool for analysing data and assisting many physics analyses at the \ac{LHC}~\mbox{\cite{plehn2022, Guest2018, karagiorgi2021, rev_ML_ad2024}}. 
\mbox{Meanwhile}, quantum computing is a fundamentally different paradigm for information processing, that is known to provide computational speed-ups over classical methods for a large class of problems~\cite{grover1996fast, shor_polynomial-time_1997,Harrow_Hassidim_Lloyd_2009, arute_quantum_2019, zhong_quantum_2020, madsen_quantum_2022, Babbush2023}. 
Furthermore, \ac{QML} has the potential to enhance traditional \ac{ML} methods~\cite{Biamonte2017, Schuld_2018_book, Benedetti2019, Havlicek2019, Schuld_2019, circuit_centric_Schuld2020} and yields various advantages in specific learning tasks~\cite{Rebentrost2014, Liu2021rigorous, Huang2021, huangQA2022, muser2023, pirnay_super-polynomial_2022, gyurik2022}. 
\mbox{Recent} studies have highlighted guarantees regarding the expressivity, generalisation power, and trainability of quantum models~\cite{Perez2020, schuld_qml_is_kernel2021, Goto2021, abbas_power_2021, Caro_2022, Jerbi2023}. 
Moreover, the efficacy of applying \ac{QML} models to \ac{HEP} data analysis is exemplified in studies for classification~\mbox{\cite{wu2021_kernel, wu2021_vqc, terashi2021, Belis2021, blance_quantum_2021, Gianelle2022}}, reconstruction~\cite{Lerjarza2022, Tuysuz_2020, magano_quantum_2022}, anomaly detection~\cite{Ngairangbam_2022, Schuhmacher23, alvi_quantum_2022,Belis:2023atb}, and Monte Carlo integration~\cite{delejarza2024loop,Agliardi_2022}. 
A summary of advancements in \ac{QML} applied to \ac{HEP} is found in Ref.~\cite{dimeglio2023quantum}.

However, for most realistic applications, the dimensionality of the dataset is usually too large to be directly processed by commonly available quantum \mbox{computers}. 
Consequently, dimensionality reduction techniques are typically employed and treated as a preprocessing step before the data is loaded into the \ac{QML} algorithm. 
\mbox{Previous} studies use manual feature selection, informed by prior knowledge about the given problem~\cite{terashi2021, Belis2021}, \mbox{feature} extraction techniques, such as the popular \ac{PCA}~\cite{wu2021_kernel, wu2021_vqc, Schuhmacher23}, or more recently, dimensionality reduction using deep learning models, e.g., simple auto-encoders~\cite{Belis2021, Belis:2023atb, ballard1987modular}. 
\mbox{However}, there is no guarantee or even incentive for the lower-dimensional representation produced by these methods to preserve original data structures that are relevant for the \ac{QML} task, e.g., binary classification. 

Crucial information required in discriminating between the classes can be lost in the dimensionality~\mbox{reduction}. 
\mbox{In~extreme} cases, the dimensionality reduction performed as a preprocessing step can render impossible the Higgs classification task to be solved by the \ac{QML} model. 
To address this challenge, we draw inspiration from the multi-task learning literature \cite{caruana1997multitask, ruder2017overview} and design a model architecture that generates lower dimensional representations which are suitable for classification by \ac{QML} models.
We call this approach \textit{guided quantum compression}. 

In contrast with conventional techniques, the dimensionality reduction performed within the guided quantum compression framework is not treated as a separate preprocessing step.
For example, the usual approach is to train an auto-encoder on the data, obtain a lower dimensional representation from its latent space, and then train a \ac{QML} algorithm on this latter \mbox{representation}~\cite{Belis2021}.
Instead, guided quantum compression \textit{simultaneously} accomplishes the dimensionality reduction and classification tasks with a single architecture. 
This way, the performance of the \ac{QML} is not limited by the arbitrary choice of the reduction method, be it manual, classical feature extraction, or deep learning. 
\begin{figure*}[thb]
    \centering
    \includegraphics[width=\textwidth]{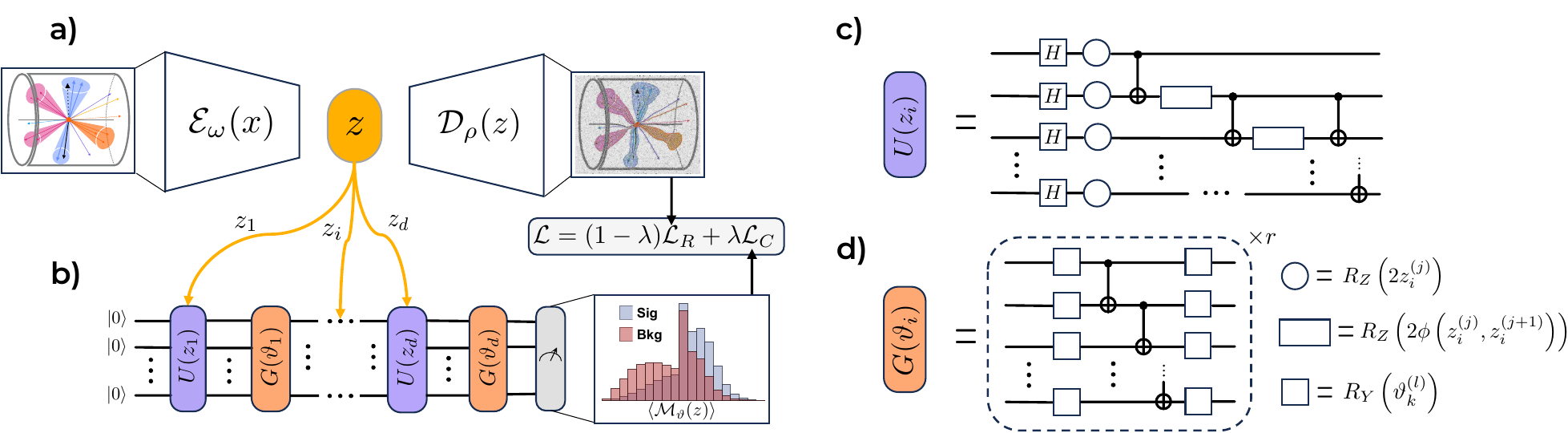}
    \caption{\textit{The GQC network.} The architecture of the Guided Quantum Compression (GQC) network is shown in \textbf{a)} and \textbf{b)}. The auto-encoder receives data from simulated LHC proton-proton collisions and produces a lower dimensional representation $z\in\mathbb{R}^\ell$ via the encoder network $\mathcal{E_\omega}$, where $\ell$ is called the latent space dimension. The decoder network $\mathcal{D_\rho}$, receives $z$ and aims to reconstruct the original data~$x$. The distinct segments $z_1,\dots,\, z_i, \dots,\, z_d$ of the latent space vector $z$ are encoded sequentially in the quantum circuit by using the feature map $U(\cdot)$; the dimension of $z_i$ is equal to the number of qubits $n$ in the circuit. The trainable gates $G(\cdot)$ are placed between the quantum encoding gates $U(\cdot)$. The output of the decoder network and quantum model are used to minimize different parts of the total loss function $\mathcal{L}$ from Eq.\,\ref{eq:gq_loss}. \textbf{c)} The data encoding circuit $U(\cdot)$ \cite{Havlicek2019} described in Sec.\,\ref{sec:vqc}. \textbf{d)}~The~variational ansatz $G(\cdot)$. The indices $j=1, \dots,\, n$ enumerate the elements of $z_i$. Moreover, the indices $l=1,\dots,\,2nr$ pertain to the trainable parameter of the corresponding $k$-th parametrised circuit block $G(\vartheta_k)$, where $r$ are the repetitions of the trainable ansatz.}
    \label{fig:model}
\end{figure*}

We demonstrate this claim by considering a realistic and complex classification task: identifying the Higgs boson in the \tth\;semi-leptonic channel for simulated proton collision data at the \ac{LHC}. 
On this data, the conventional reduction methods mentioned earlier fail relative to the guided quantum compression method, i.e., independently compressing the dataset before training the \ac{QML} model leads to poor classification performance. 
In contrast, the guided quantum compression method is able to solve the classification problem, reaching competitive accuracy with state-of-the-art classical methods~\cite{reissel21}. 
Furthermore, we observe an improved performance of our algorithm compared to the classical benchmark when using only the particle kinematics, suggesting that for \ac{QML} one should use features representative of the quantum process that generated the data~\cite{top_correlations_CMS2019, Kubler2021, Schuhmacher23, Belis:2023atb}.

\section{Models}\label{sec:models}
The guided quantum compression network shown in Fig.\,\ref{fig:model} is comprised of an auto-encoder~(Fig.\,\ref{fig:model}a) and a variational quantum circuit (Fig.~\ref{fig:model}b) that are coupled. \mbox{In the following}, we describe these two elements and how they are simultaneously trained.

\subsection{The Auto-encoder}\label{sec:AE}

The \ac{AE} is a machine learning model that has   been used for decades across the historical landscape of neural networks \cite{lecun1987phd, ballard1987modular, hinton1993autoencoders}. The most commonly used type of \ac{AE} consists of two feed-forward neural networks: the encoder $\mathcal{E}_\omega$ and the decoder $\mathcal{D}_\rho$. The~encoder $\mathcal{E}_\omega$ maps the input feature space $x$ to a \textit{latent space} $z$ of lower dimension $\ell$. Conversely, the goal of the decoder $\mathcal{D}_\rho$ is to reconstruct the input $x$ from~$z$. A~schematic of an \ac{AE} neural network is shown in Fig.\,\ref{fig:model}a. The objective of the \ac{AE} training is to minimise the difference between the input data and the reconstructed~data; this difference can be quantified by \mbox{various functions}. There exist many types of \ac{AE}s, based on how this difference is quantified or on architectural extensions \cite{Kingma:2013hel, Deja:2020tag, joo2020dirichlet, pan2018adversarially}. The \ac{MSE} is the standard function used to quantify the difference between the input data $x$ and its reconstructed counterpart:
\begin{equation}
\label{eq:vanillaloss}
\mathcal{L}_R = \frac{1}{M}\sum^M_{m=1}\left[x_m - \mathcal{D}_\rho\circ\mathcal{E}_\omega(x_m)\right]^2,    
\end{equation}
where $\mathcal{D_\rho}$ is the decoder network with weights $\rho$, $\mathcal{E}_\omega$ is the encoder network with weights $\omega$, and $M$ is the size of the training dataset. The conventional \ac{AE} learns as any other feed-forward neural network, with one subtlety: the reconstruction loss is not only propagated through the decoder network, but through the encoder as well. Therefore, the latent space and the reconstructed data evolve simultaneously as the \ac{AE} model is learning.

\subsection{The Variational Quantum Circuit Classifier}\label{sec:vqc}
The \ac{VQC}~\cite{Benedetti2019, Mitarai2018} is a \ac{QML} model based on parametrized quantum circuits that are trained variationally to undertake tasks such as classification~\cite{blance_quantum_2021, Belis2021, wu2021_vqc, terashi2021}, regression~\cite{Perez-Salinas2020}, and generative modelling~\cite{Chang2021, Kiss2022, Delgado2022, Bravo2022}. The classifier output, in the \ac{VQC} implementation, is the expectation of an observable $\mathcal{M}$. This expectation is interpreted as the likelihood of the input sample to belong in a certain data class, e.g., the sample contains a Higgs boson or not, and is extracted from the quantum computer via doing measurements. Specifically, the model output $f_\vartheta(z)$, for a given input data vector $z$ and gate parameters $\vartheta$, is defined as
\begin{equation}
    f_\vartheta(z)=\Braket{0|\mathcal{U}_\vartheta^\dagger(z)\mathcal{M}\,\mathcal{U}_\vartheta\left(z\right)|0}=\Braket{\mathcal{M}_\vartheta(z)},
    \label{eq:vqc_def}
\end{equation}
where $\mathcal{U}_\vartheta(z)$ is the whole quantum circuit of the model, $\mathcal{M}$ is the observable whose expectation value we measure, and $\ket{0}=\ket{0}^{\otimes n}$ the initial $n$ qubit state. Furthermore, the label predicted by the \ac{VQC}, $\hat{y}\in \{0,1\}$, is given by
\begin{equation}
    \hat{y} = \frac{\text{sign}\left[\braket{\mathcal{M}_\vartheta(z)}\right]+1}{2},
    \label{eq:pred}
\end{equation}
where one can assume that $\braket{\mathcal{M}_\vartheta(z)}\in [-1, 1]$ without any loss of generality.

We design a \ac{VQC} architecture as shown in Fig.\,\ref{fig:model}b. The input data vector $z$ is split into $d$ sub-vectors $z=(z_1, z_2,\dots ,z_d )$, each $z_i$ being of dimensionality $n$. \mbox{Furthermore}, each sub-vector is encoded into the quantum circuit sequentially by using the feature map $U(\cdot)$; between each $U(\cdot)$ there is a set of trainable gates $G(\cdot)$. Hence, the whole quantum circuit of the model is 
\begin{equation}
    \mathcal{U}_\vartheta(z) = \prod_{k=1}^{d}G(\vartheta_k)U(z_k).
\end{equation}

The proposed architecture design is theoretically motivated by how the \ac{VQC} model expressivity increases with the circuit depth~\cite{Perez2020, Schuld_Sweke_Meyer_2021}. Furthermore, through our encoding strategy, we are able to tune the number of qubits and the number of gate operations in the \ac{VQC} circuit. This way, we ensure $d=\ell/n$, where $d$ is the number of segments of the latent vector $z$ and $\ell$ is the dimensionality of the latent space produced by the auto-encoder.


Here, we set the observable $\mathcal{M}\equiv\sigma_z$, where $\sigma_z$ is the Pauli-Z operator acting on the first qubit. The data encoding $U(\cdot)$ used herein is the feature map from Ref.~\cite{Havlicek2019}. This consists of $R_z$ rotation gates, which encode one data feature per qubit, and nearest-neighbor entanglement between these qubits, as presented in Fig.\,\ref{fig:model}c. The data encoding blocks include interactions between the features: 
\begin{equation}
\phi(z_i^{(j)},z_i^{(j+1)})=\prod_{j=1}^n\left(\pi-z_i^{(j)}\right)\left(\pi-z_i^{(j+1)}\right),
\end{equation}
where the symbols are described in Fig.\,\ref{fig:model}c and Fig.\,\ref{fig:model}d. 
A~single repetition of the variational ansatz applies $R_Y$ rotations on each qubit and nearest-neighbor entanglement via the CNOT (CX) gates, as in Fig.\,\ref{fig:model}d. 

We choose these general purpose data encoding and trainable circuits to highlight the effectiveness of the GQC network in solving classification tasks that conventional methods would struggle with. Hence, we do not focus on an extensive search for a quantum circuit that would yield the best classifier performance; our results do not depend on a specific \ac{VQC} architecture choice. 

All classifiers in this work are trained to minimise the binary cross entropy loss function: 
\begin{equation}
    \mathcal{L}_C = -\frac{1}{M} \sum_{m=1}^{M} \left[ y_m \log(\hat{y}_m) + (1 - y_m) \log(1 - \hat{y}_m) \right], 
    \label{eq:loss_bce}
\end{equation}
where $M$ is the number of data points, e.g., in a batch, $y_m$ is the true label of the data point $m$, and $\hat{y}_m$ is its predicted label, as in Eq.\,\ref{eq:pred}.
This includes the \ac{VQC} algorithm presented in Sec.\,\ref{sec:2step}, any traditional methods, defined in Sec.\,\ref{sec:classical}, and the \ac{VQC} part of the guided quantum compression paradigm, introduced in Sec.\,\ref{sec:gqc_training}.
All of these are precisely defined in the following section.
This scoring rule dates back to 1952 \cite{good1952rational}, albeit introduced in a different context, and is the most popular loss function used for \ac{ML} classification tasks. 

\newpage
\section{Training Paradigms}\label{sec:paradigms}

Different training methods are investigated to solve the binary classification problem of the \tth\;dataset~\cite{tthbb_data}, in order to contrast and exemplify the dimensionality reduction advantages of the guided quantum compression algorithm introduced in this work. 
Specifically, we study two processes occurring in proton-proton collision events at the LHC: the \textit{signal} \tth\;process, in which a Higgs boson is produced, and the \textit{background} \ttbb\;process, where it is not produced. 
These two processes lead to the same final-state particles, generating similar high-dimensional signatures in the~\mbox{detector}. 
The~task of distinguishing these two processes is challenging~\cite{Belis2021, tthbb_data}.
However, accurate classification of such events is crucial for enhancing the sensitivity of LHC experiments, enabling study Higgs boson processes with high \mbox{precision}.
\mbox{Furthermore}, we assume that this dataset is representative of common high-dimensional datasets in which the probability distributions of the features for the two classes have substantial overlap.
For these reasons, this dataset and the corresponding classification task is well-suited for \ac{ML} and \ac{QML} methods.
It enables us to demonstrate that the proposed guided quantum compression paradigm generates a better latent space compared to conventional techniques in a challenging real-world problem, important to fundamental physics research. 

To this end, we use three distinct training paradigms. First, the so-called classical approach, which serves as our baseline benchmark. Second, the 2Step method, which represents the usual way to perform dimensionality reduction when using \ac{QML} for classifying complex data. Finally, the GQC paradigm developed in this work, which performs the dimensionality reduction and classification objectives at the same time.

Note, the three aforementioned strategies pertain to different computational requirements: it is less computationally intensive to train and robustly optimise the hyperparameters of a classical algorithm compared to a quantum one. 
Thus, while we are able to find the optimal values for the parameters of the classical network, this is not technically achievable for the quantum models presented in this work. 
The amount of compute time required to find the optimal hyperparameters for the quantum circuit is simply unfeasible. 
Furthermore, a vast amount of computational resources (cf.\,App.\,\ref{app:hyper}) are also required to find the best quantum hyperparameter combination, including the best number of ansatz repetitions, the best number of qubits, or the best learning rate.

We present the classical and 2Step methods since they contrast and highlight the advantages our method brings in performing the needed dimensionality \mbox{reduction}.
The~classical baseline is used to disambiguate the performance of the quantum algorithm and classical component of the guided quantum compression paradigm.
\mbox{Meanwhile}, the 2Step method is used to contrast the guided quantum compression with the conventional way of applying \ac{QML} to our high-dimensional data.

\newpage
\subsection{Classical}\label{sec:classical}
A fully-connected feed-forward network is our classical benchmark. 
This network is trained to minimise the binary cross entropy loss, in Eq.\,\ref{eq:loss_bce}, via stochastic gradient descent. 
In~essence, the classical paradigm represents the most conventional way to address our classification task. 

The hyperparameters of the feed-forward network, such as the learning rate, are optimized via a grid~search. This is the only type of model and paradigm for which an exhaustive hyperparameter optimisation is feasible; see App.~\ref{app:hyper} for more details.

\subsection{2Step}\label{sec:2step}
\begin{figure*}[thb]
    \centering
    \includegraphics[width=0.48\textwidth]{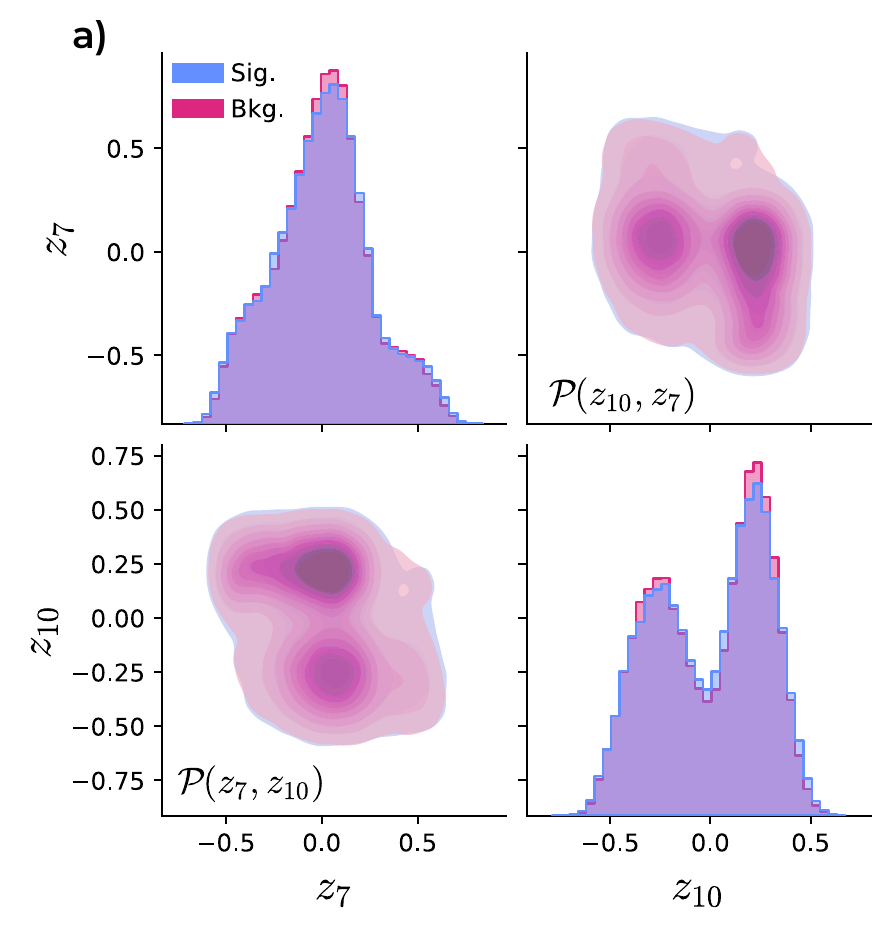}\hspace{0.5cm}
    ~\includegraphics[width=0.48\textwidth]{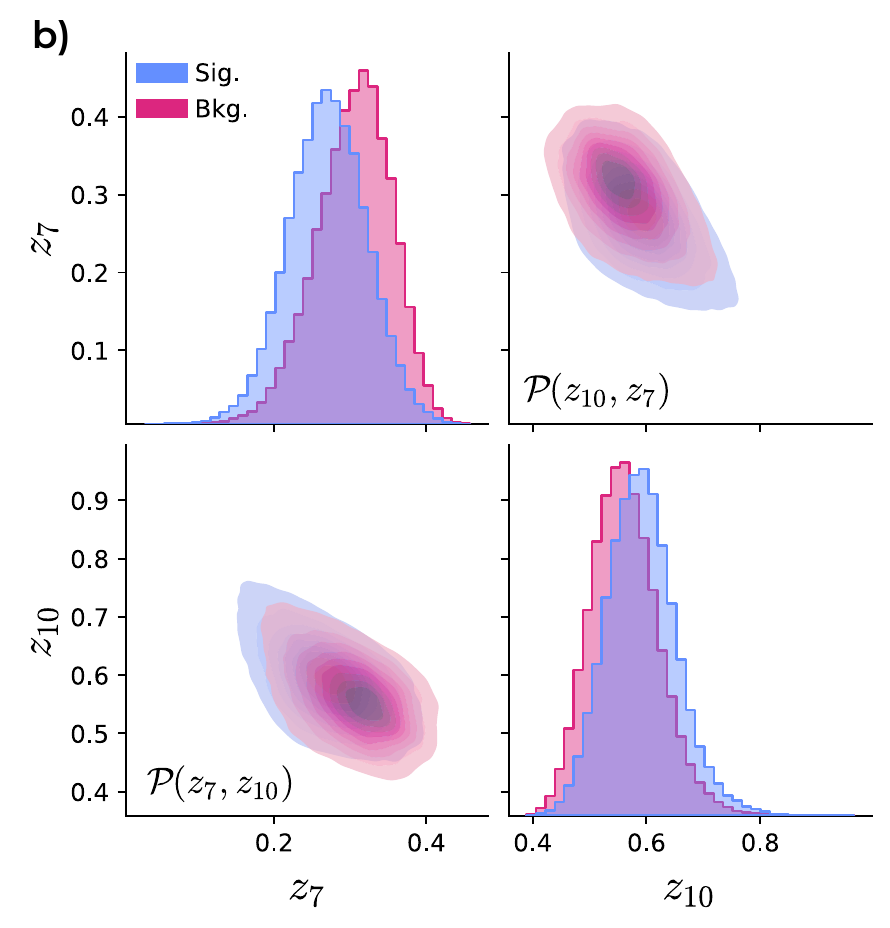}
    \caption{\textit{Latent representation.} The one- and two-dimensional projections of the \tth\;dataset latent space $z\in\mathbb{R}^\ell$ generated by \textbf{a)} the 2Step training paradigm and \textbf{b)} GQC model. The probability distributions of the latent features $z_7$ and $z_{10}$ are shown in the histogram plots. The joint two-dimensional probability distributions $\mathcal{P}(z_7, z_{10})$ are displayed in the density plots. Notice~that the latent space separation of signal and background is better in the GQC algorithm; furthermore, the GQC latent distributions are more regularly shaped. The latent features $z_7$ and $z_{10}$ are arbitrarily chosen to show the structure of the latent vector $z$ in one or two dimensions. These joint distributions are symmetric: $\mathcal{P}(z_7,z_{10})=\mathcal{P}(z_{10},z_{7}).$}
    \label{fig:latent_rep}
\end{figure*}
In the 2Step paradigm, the dimensionality reduction algorithm, namely the \ac{AE} from Sec.~\ref{sec:AE}, is trained independently from the \ac{VQC} classifier presented in Sec.~\ref{sec:vqc}. As~the~name suggests, the classification task is performed in two separate steps. First, the hyperparameters of the \ac{AE} are optimised and the resulting architecture is trained with the goal of minimising the \ac{MSE} loss between the input data and the output of the decoder. Secondly, the \ac{VQC} is trained by using the latent space representation $z$ of each sample that the \ac{AE} produces. This way, the \ac{VQC} acts on a lower dimensional input and consequently its quantum circuit can be smaller. As mentioned in Sec.\,\ref{sec:intro}, the dimensionality reduction allows us to employ a reasonable amount of resources in simulating quantum circuits in classical devices and in implementing our quantum models in currently available quantum computers. Hitherto, we outlined the most common way in which \ac{AE}s are used for dimensionality reduction in the context of a classification task \cite{wang2016auto}; in the following, a paradigm which integrates the two steps is introduced.

\subsection{Guided Quantum Compression}\label{sec:gqc_training}
Our strategy aims to address the problem of generating a low-dimensional representation that ensures the discrimination of the classes by the quantum model. The GQC network implements a trainable data encoding map $\mathcal{U}_\vartheta\left(\mathcal{E}_\omega(x)\right)$. The lower-dimensional representation learned by the encoder network is \textit{guided} by the quantum classification algorithm.

\begin{figure*}[thb]
    \centering
    \includegraphics[width=0.5\textwidth]{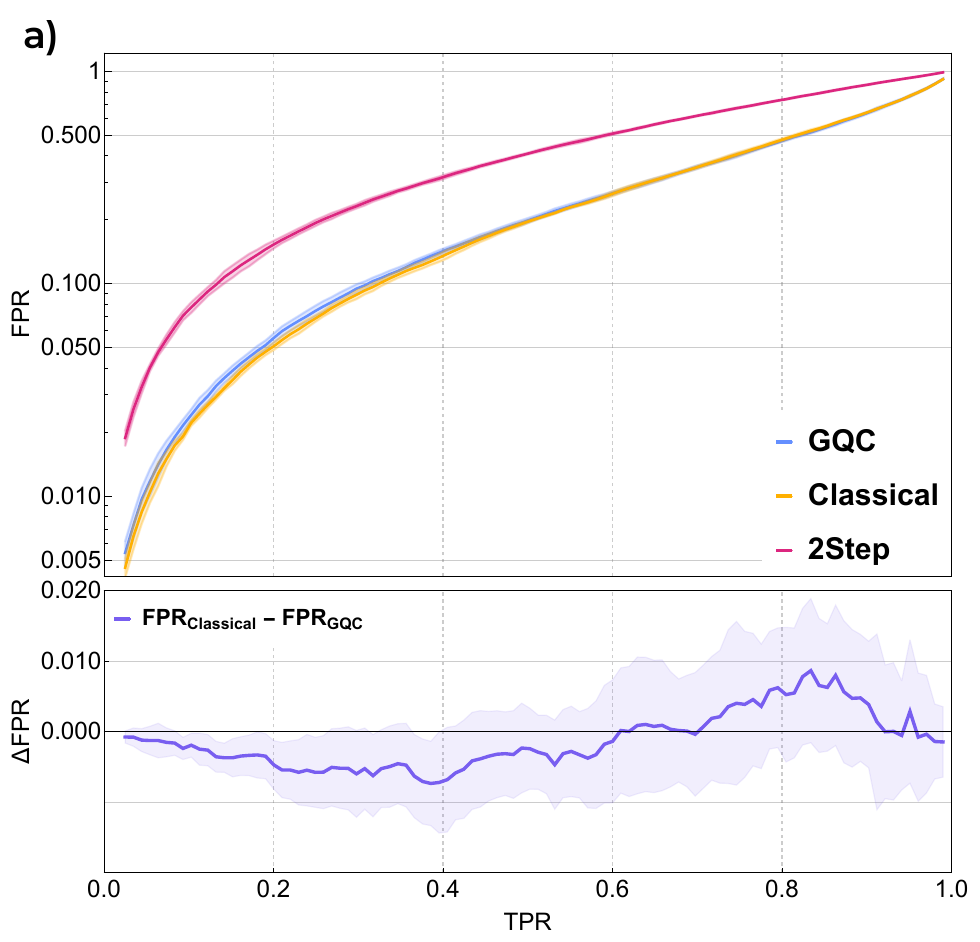}
    ~\includegraphics[width=0.5\textwidth]{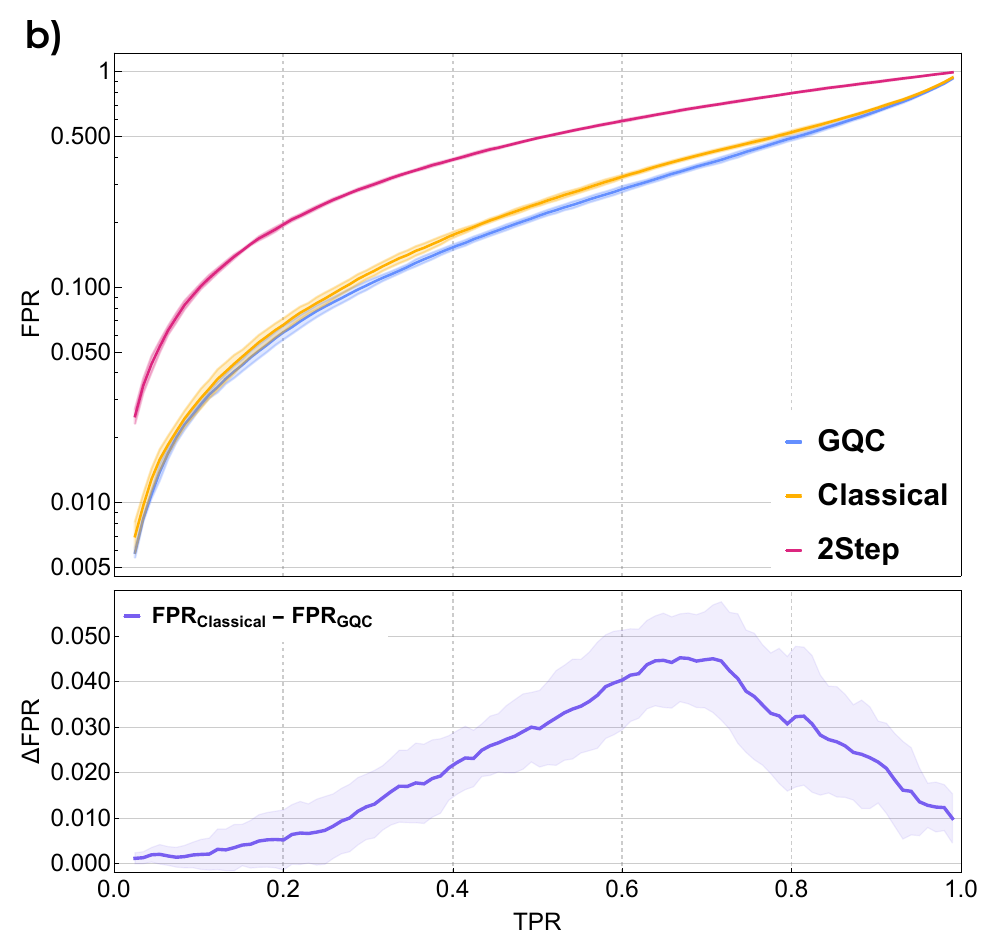}
    \caption{\textit{Receiver Operating Curves.} The ROC curves of the models with \textbf{a)} the btag features and \textbf{b)} without the btag features included in the training data. The 2Step training procedure yields the worst performance. At the bottom panel of each plot the difference between the GQC ROC and the classical ROC is displayed. When the btag is absent from the dataset, the GQC model outperforms the classical benchmark in the TPR range of $0.4$ to $0.9$, as shown in the lower panel of \textbf{b)}. }
    \label{fig:rocs}
\end{figure*}

The guidance is performed during the training of the GQC network by coupling the auto-encoder and \ac{VQC} models through the following loss function:
\begin{equation}
    \mathcal{L}=(1-\lambda)\mathcal{L}_R + \lambda\mathcal{L}_C,
\label{eq:gq_loss}
\end{equation}
where $\mathcal{L}_R$ is the MSE loss as defined in Eq.\,\ref{eq:vanillaloss}, $\mathcal{L}_C$ is the loss of the \ac{VQC} classifier defined in Eq.\,\ref{eq:loss_bce}, and $\lambda\in(0,1)$ is the hyperparameter that specifies the coupling between the reconstruction and classification optimisation tasks.
The $\lambda$ is fine-tuned separately to maximise the classification accuracy on the validation dataset, see App.~\ref{app:hyper}.

The simultaneous learning of the two objectives improves the generalisation power of the classifier~\cite{Lei2018, vafaeikia2020}. 
This synergy arises from the regularisation imposed through Eq.\,\ref{eq:gq_loss} of the quantum classifier. 
Namely, the classifier performance is enhanced by the additional task of reconstructing data from the latent space~\cite{vafaeikia2020}.

In our algorithm, we train the GQC network via stochastic gradient descent with the Adam optimizer~\cite{kingma2017adam}. The gradients of the classical parts of the model are computed using backpropagation, while the quantum circuit gradients are computed via the adjoint differentiation method~\cite{jones2020}, for efficient training of the model on classical processors used to simulate the quantum software. 
The~proposed GQC architecture can also be trained on quantum hardware using the parameter-shift rule~\cite{schuld19_paramshift} \mbox{instead} of the adjoint differentiation method used here. 
The hyperparameters of the GQC network are tuned using a sequential grid search for each hyperparameter while keeping all the rest fixed. For more details on the hyperparameter tuning procedure, see App.\,\ref{app:hyper}.


\section{Results}\label{sec:results}

\begin{table*}[t]
\begin{ruledtabular}
\centering
\begin{tabular}{l c c c c} 
Model &  AUC w/ b-tag &  AUC w/o b-tag &  FPR\textsuperscript{-1} w/ b-tag & FPR\textsuperscript{-1} w/o b-tag  \\
\midrule
\texttt{GQC} &  $0.733 \pm 0.003$ &  $\mathbf{0.720 \pm 0.005}$&  $2.134 \pm 0.028$& $2.045 \pm 0.049$\\ 
\texttt{2Step} &  $0.561 \pm 0.003$&  $\mathbf{0.508 \pm 0.002}$&  $1.263 \pm 0.004$& $1.368 \pm 0.007$         \\ 
\texttt{Classical} &  $0.734 \pm 0.002$&  $\mathbf{0.699 \pm 0.004}$&  $2.107 \pm 0.029$& $1.921 \pm 0.035$     \\ 
\end{tabular}
\caption{\textit{Model Performances.} The first two columns present the Area Under the Curve (AUC) values of the three training paradigms with and without the b-tag variable in the training data, respectively. The last two columns show the inverse FPR at $\mathrm{TPR}=0.8$. }
    \label{tab:results}
\end{ruledtabular}
\end{table*}

The classification performance of the three training paradigms described in Sec.\,\ref{sec:paradigms} is benchmarked using the same simulated dataset consisting of \tth\;(signal) and \ttbb\;(background) events~\cite{tthbb_data, Sirunyan2019, ATLAS:2020syy}. 
For all models except the \ac{AE} used in the 2Step paradigm, the training data consists of 20,000 samples; meanwhile, the \ac{AE} uses $1.44\times10^6$ samples for training. 
The test dataset consists of 5 k-folds, with 20,000 samples for each fold. 
The~\mbox{number} of samples is balanced across the two classes for all datasets. 
Additionally, a validation dataset of 1,500 samples is used during training to monitor for overfitting and to fix the hyperparameters of the models.
Specifically, at the end of each training epoch, the validation loss is computed on the the validation dataset for each model: the classification loss in Eq.~\ref{eq:loss_bce} for the classifier models, the reconstruction loss in Eq.~\ref{eq:vanillaloss} for the \ac{AE}, or the combined loss in Eq.~\ref{eq:gq_loss} for GQC model. 
In each case, the model that yields the minimum validation loss is selected after the training. 

The initial dimensionality of each data sample is 67 (60), if high-level, classically preprocessed, features pertaining to the collision event are included (excluded). 
This is always reduced to a final dimensionality of 16 by the \ac{AE}, except in the classical training workflow from Sec.\,\ref{sec:classical}, which does not use dimensionality reduction.
The initial 60 features are the kinematic variables of the particles included in the collision event, e.g., their energy, angle of incidence on the detector, and so on~\cite{tthbb_data}. 
These kinematic variables describe the quantum interaction of colliding particles and can be calculated from first-principles within the framework of quantum field theory.
\mbox{Furthermore}, the additional 7 features that introduce high-level properties of the collision event are the so called \textit{btag} variables. 
These determine the likelihood of a certain particle, the $b$ quark, being produced in the~event. 
These likelihoods are conventionally obtained using classical ML models when producing this dataset. 

The dataset is preprocessed using physical criteria before the training and testing of the investigated models. 
The exact requirements can be found in Refs.~\cite{tthbb_data, Belis2021}.
These preprocessing steps accommodate for the geometric acceptance of the detector used to record the proton collision data and ensure that each collision event has the desired characteristics for the study, e.g., at least one detected lepton. 
Subsequently, we normalise all data features to the range $[0, 1]$. 

The lower dimensional latent spaces produced through the conventional 2Step method and the GQC paradigm are shown in Fig.\,\ref{fig:latent_rep}.
As clearly visible in this figure, the GQC network learns a better separation between the two data classes in its latent space. 
Hence, it is favorable to use the GQC paradigm when performing dimensionality reduction on the input data.
In App.\,\ref{app:kl}, we quantify the class separation and the GQC improvement using feature-wise Kullback-Leibler divergences~\cite{Kullback_Leibler1951}.

The \ac{ROC} produced through each paradigm from Sec.~\ref{sec:paradigms} is shown in~Fig.\,\ref{fig:rocs}. 
The 5 k-folds of the test data are used to compute these \ac{ROC} curves and their corresponding \mbox{uncertainties}.
The~uncertainty is represented as error bands with a width of one standard deviation around the ROC curves. 
The conventional 2Step method performs the worst. 
Meanwhile, the classical and the GQC methods yield a similar classification accuracy when the btag is included in the dataset, c.f. Fig.\,\ref{fig:rocs}b; excluding the btag, i.e., keeping only the kinematic variables of the particles, leads to the GQC significantly outperforming the classical approach. 
The advantage in classification performance appears in the relevant range of True Positive Rate (TPR) between 0.4 and 0.9, which is a typical choice for physics analyses at the \ac{LHC}. 
The results in Fig.\,\ref{fig:rocs} are condensed in Tab.\,\ref{tab:results}, where summary performance metrics derived from these \ac{ROC}s are shown.
The GQC latent space allows for a competitive performance with the classical benchmark.

\section{Conclusions}\label{sec:conclusions}

The choice of the compression method can have a significant impact on the classifier performance. We show that for the \tth\;classification task, applying dimensionality reduction as a preprocessing step renders the problem impenetrable for a currently applicable VQC classifier algorithm, as highlighted in Fig.\,\ref{fig:rocs}~and~Tab.\,\ref{tab:results}. 
In~contrast, when an identical \ac{VQC} is used as part of the GQC training paradigm, its classification performance drastically increases.
Furthermore, the GQC model achieves a consistent enhanced classification accuracy without an increase in its training time complexity compared to the 2Step approach; it inherits its training time complexity from its components: the autoencoder and the \ac{VQC}.
Thus, integrating the dimensionality reduction and classification tasks is shown to provide a better low dimensional latent space for our problem.

Additionally, the hyperparameter optimisation procedure employed for each network in this work leads to a potential bias towards the classical paradigm when comparing the model performances: it~is~computationally feasible to perform an extensive hyperparameter optimisation for the classical algorithm, while this is currently not the case for the GQC model. 
Thus, the hyperparameters of the presented networks that include a QML element are only approximately optimal while the hyperparameters of the classical algorithm are the best they can possibly be. 
Nevertheless, the conclusions we draw from our results are not affected by these limitations.

For simple datasets, dimensionality reduction and classification tasks can potentially be decoupled, i.e., they can be treated as independent problems~\cite{wu2021_kernel, wu2021_vqc, Belis2021, Schuhmacher23}. 
However, for realistic datasets it is possible that such a separate treatment yields worse results. 
The dimensionality reduction algorithm can obfuscate the class structure of the original problem, apparent in our results. 
The~GQC encoding strategy ensures that the \ac{VQC} is both expressive and flexible for choosing the desired circuit width and depth, while retaining its overall \mbox{accuracy}.
The training paradigms presented in this work show the benefits of carefully constructing hybrid \ac{QML} \mbox{models}. 
These advantages stem from aligning the goals of the dimensionality reduction algorithm, typically chosen a priori and arbitrarily, with that of the quantum model. Hence, the use of classical methods for dimensionality reduction can facilitate current quantum computing applicability in realistic settings.

As seen in Fig.~\ref{fig:rocs}a, in the worst case the performance of the GQC network is equivalent to that of the classical benchmark. The~GQC network outperforms both the classical model and the quantum model in the 2Step approach when the highly processed btag feature is not used in the~\mbox{training}, shown in Fig.\,\ref{fig:rocs}b.
Moreover, it is competitive with state-of-the-art results on this data subset~\cite{reissel21}.
This suggests that for improved QML performance on particle physics data, features that are representative of the quantum process that generated it~\cite{top_correlations_CMS2019} are preferred, e.g., the angular distribution of particles~\mbox{\cite{Schuhmacher23, Belis:2023atb}}. 
\mbox{Investigating} this in more detail is left for future work.

\vfill\null

\section*{Data availability}\label{sec:data}
The datasets used for this study are publicly available on Zenodo~\cite{tthbb_data}.

\section*{Code availability}\label{sec:code}
The code developed for this paper is available publicly in the GitHub repository: \url{https://github.com/CERN-IT-INNOVATION/gqc}. \mbox{A~combination} of PyTorch and PennyLane~\cite{bergholm2022pennylane} were used to implement and train the GQC, 2Step, and the classical model architectures. The website in Ref.\,\cite{deepfried} was used for Fig.\,\ref{fig:model}a to distort the reconstructed high-energy physics event schematic.

\section*{Acknowledgements}

V.B. is supported by an ETH Research Grant (grant no.~ETH C-04 21-2). P.O. is supported by Swiss National Science Foundation Grant No.~PZ00P2\_201594. The authors would like to thank Elías F. Combarro for the helpful discussions. M.G. and S.V. are supported by CERN through the Quantum Technology Initiative. F.R. acknowledges financial support by the Swiss National Science Foundation (Ambizione grant no. PZ00P2$\_$186040).
\vfill

\bibliography{main}

\onecolumngrid
\appendix
\newpage

\section{Model Hyperparameter Selection}\label{app:hyper}
The classical autoencoder used in this study for the 2Step paradigm has an encoder network consisting of six layers with nodes [67, 64, 44, 32, 24, 16], respectively. The decoder network mirrors the architecture of the encoder: [16, 24, 32, 44, 64, 67]. The ReLU activation function is applied to all layers except for the layer corresponding to the latent space, i.e. the one with 16 nodes, and the output layer, i.e., the last layer of the decoder with 67 nodes. For the latent and output layers both the sigmoid and tanh activation functions have been assessed providing similar results. 

The hyperparameters of all the classical models are found through exhaustive grid search. Specifically, we repeat the training and k-fold testing of each model, as described in Sec.~\ref{sec:results}, for each hyperparameter combination. The considered hyperparameter combination is $\{32,\, 64,\, 128,\, 256,\, 512,\, 1024,\, 2048\}$ for the batch size and $\{10^{-3},\,10^{-2},\,10^{-1}\,\}$ for the learning rate. 

For the \ac{AE}, we firstly identify that a batch size of 128 and a learning rate of $10^{-3}$ yields the best performance. With the batch size fixed at 128, we subsequently perform a finer hyperparameter tuning for the learning rate in the neighborhood of $[10^{-3}, 10^{-2}$], using the Optuna hyperparameter optimisation framework~\cite{DBLP:journals/corr/abs-1907-10902}. Following this procedure, we find an optimal learning rate of 0.0012.

For the classical benchmark, we evaluate the performance of a fully connected feed-forward network. This network has the same architecture as the encoder network of the \ac{AE} model, with the addition of a single-node layer with a sigmoid activation function serving as the output of the classifier. For studies excluding the btag variable, the model architectures stay the same, except that the input layer is changed to 60 nodes. Additionally, for the \ac{AE}, the output layer is also set to 60. Furthermore, we investigate a shallower architecture with three layers in total [67, 16, 1] that yielded similar classification performance. Hence, fos simplicity, we choose the shallower architecture. To maintain consistency with the classical benchmark, the classical components of the hybrid GQC network (cf.~Fig.~\ref{fig:model}), $\mathcal{E}_\omega$ and $\mathcal{D}_\rho$, also have the shallow architectures of [67, 16] and [16, 67], respectively.

In the following, the sequential grid search we employed to tune the hyperparamters of the GQC network is described; in bold we present the hyperparameter values found to be optimal following our procedure. For our simulations and numerical studies, we used specialized nodes at the CERN and Paul Scherrer Institut PSI (PSI) computing clusters. 
At each step of the the sequential grid search the models are retrained and evaluated using k-fold testing, as discussed in Sec.~\ref{sec:results}. Firstly, we choose the best performing batch size out of $\{128,\,256,\,512,\, \textbf{1024},\, 2048\}$, afterwards we do a scan on the learning rate $\{10^{-3},\,\mathbf{10^{-2}},\,10^{-1}\}$, then we optimize for the repetitions of the trainable gates (cf.\,Fig.\,\ref{fig:model}) of the \ac{VQC} with $r\in\{\textbf{2},\,4,\,8\}$. Lastly, we fix all the above hyperparameters and optimize the coupling $\lambda$ (see Eq.\,\ref{eq:gq_loss}) in the range $[0.3,\,0.9]$ with a step of $0.1$ finding the best performing value at $\lambda=\mathbf{0.7}$. The $\lambda$ is fine-tuned to maximise the classification accuracy on the validation dataset.

An exhaustive grid search, as the one performed for the classical model (cf.\,Sec.~\ref{sec:results}), would require assessing all possible combinations of the above hyperparameters leading to a drastic increase of compute time. Specifically, it would increase the current quantum simulation runtime, which is at the order of days, to multiple weeks.

\section{Quantifying the class separation in the latent distributions}
\label{app:kl}

The Kullback-Leibler divergence (KLD), also called relative entropy, is a measure that quantifies the difference between a probability distribution $P$ and a reference probability distribution $Q$~\cite{Kullback_Leibler1951}. Specifically, given a continuous random variable $x\in\mathcal{X}$, over a sampling space $\mathcal{X}$, the KLD is defined as
\begin{equation}
D_{\text{KL}}(P\, ||\, Q) = \int_{\mathcal{X}} P(x) \log \left( \frac{P(x)}{Q(x)} \right) dx.
\end{equation}
The KLD approaches zero as the distributions become more similar.
In general, one does not have access to an analytical expression for $P$ and $Q$ but only to finite samples from these distributions . In such cases, the KLD can be computed by constructing histograms from the $P$ and $Q$ samples,
\begin{equation}
D_{\text{KL}}(P\,||\,Q) = \sum_{i=k}^{N} P_k \log\left(\frac{P_k}{Q_k}\right), \label{eq:kl_bins}
\end{equation}
where $N$ is the number of bins, and $P_k$ and $Q_k$ represent the probabilities of observing values that fall into the $k$-th bin.

In our work, we are interested in quantifying the separation between the background $\mathcal{B}$ and signal $\mathcal{S}$ distributions in the latent space produced by the conventional 2Step method and the proposed GQC network, as seen in Fig.\,$\ref{fig:latent_rep}$ and discussed in Sec.\,\ref{sec:paradigms}. Specifically, using Eq.~\ref{eq:kl_bins}, we compute the difference between the background $\mathcal{B}(z_i)$ and signal $\mathcal{S}(z_i)$ latent distributions for each latent feature $z_i$, where $i=0,1,\dots, 15$ enumerates here the latent features. The obtained values are presented in Tab.~\ref{tab:kl_latent}. By computing the average ratio of KL divergences $D_{\text{KL}}^{\text{(GQC)}}$ and $D_{\text{KL}}^{\text{(2Step)}}$, obtained from the GQC network and 2Step latent distributions, respectively, over all features we arrive at
\begin{equation}
    \mathcal{R}=\frac{1}{16}\sum_{i=0}^{15}\frac{D_{\text{KL}}^{\text{(GQC)}}\left(\mathcal{S}(z_i)\,||\,\mathcal{B}(z_i)\right)}{D_{\text{KL}}^{\text{(2Step)}}\left(\mathcal{S}(z_i)\,||\,\mathcal{B}(z_i)\right)}\approx 79.58.
\end{equation}
Hence, we observe a significant improvement by a factor of 79.58 with the GQC over the 2Step method in terms of signal and background separation in the latent representations, as quantified by the KL divergence. This in turn also leads to higher classification accuracy for the GQC network, as presented in Sec.\,\ref{sec:results}.

\begin{table}[t] 
\begin{ruledtabular}
\centering
\begin{tabular}{l c c c c c c c c c c c c c c c c} 
   &  $z_0$ & $z_1$ & $z_2$ & $z_3$ & $z_4$ & $z_5$ & $z_6$ & $z_7$ & $z_8$ & $z_9$ & $z_{10}$ & $z_{11}$ & $z_{12}$ & $z_{13}$ & $z_{14}$ & $z_{15}$ \\
\midrule
$D_{\text{KL}}^{\text{(2Step)}}\left(\mathcal{S}(z_i)\,||\,\mathcal{B}(z_i)\right)$ &  $0.16$ &  $0.09$ & $0.31$ & $0.16$ & $0.34$ & $0.23$ & $0.12$ & $0.05$ & $0.07$ & $0.04$ & $0.07$ & $0.10$ & $0.03$ & $1.81$ & $0.08$ & $1.42$ \\ \\
$D_{\text{KL}}^{\text{(GQC)}}\left(\mathcal{S}(z_i)\,||\,\mathcal{B}(z_i)\right)$ &  $3.53$ &  $3.84$ & $17.60$ & $6.25$ & $2.17$ & $2.77$ & $3.12$ & $23.40$ & $7.06$ & $6.06$ & $7.23$ & $5.89$ & $4.94$ & $2.73$ & $0.41$ & $2.29$ \\ 
\end{tabular}
\textbf{}\caption{Quantifying the separation between the background $\mathcal{B}(z_i)$ and signal $\mathcal{S}(z_i)$ distributions in the latent space produced by the 2Step and GQC methods, respectively, using the KL divergence $D_{\text{KL}}$. The presented values correspond to each latent space feature $z_i$, where $i=0,1,\dots,15$.}
    \label{tab:kl_latent}
\end{ruledtabular}
\end{table}

\end{document}